\documentclass{ws-ijmpa}
\begin{document}

\def\cM{{\cal M}}
\def\cO{{\cal O}}
\def\cK{{\cal K}}
\def\cS{{\cal S}}
\newcommand{\mh}{m_h}
\newcommand{\mw}{m_W}
\newcommand{\mz}{m_Z}
\newcommand{\mt}{m_t}
\newcommand{\mb}{m_b}
\def\lsim{\mathrel{\raise.3ex\hbox{$<$\kern-.75em\lower1ex\hbox{$\sim$}}}}
\def\gsim{\mathrel{\raise.3ex\hbox{$>$\kern-.75em\lower1ex\hbox{$\sim$}}}}
\def\ga{\mathrel{\raise.3ex\hbox{$>$\kern-.75em\lower1ex\hbox{$\sim$}}}}
\def\la{\mathrel{\raise.3ex\hbox{$<$\kern-.75em\lower1ex\hbox{$\sim$}}}}

\newcommand{\non}{\nonumber}

\markboth{Mohamed Chabab} {On the implications of a dilaton in
gauge theory}

\catchline{}{}{}{}{}

\title{ON THE IMPLICATIONS OF A DILATON IN GAUGE THEORY}

\author{MOHAMED CHABAB }
\address{LPHEA, Physics Department, Faculty of Sciences -
Semlalia\\
 Cadi-Ayyad University, P.O. Box 2390\\
 Marrakech 40000, Morocco.\\
 mchabab@ucam.ac.ma}

\maketitle

\begin{abstract}
Some recent work on the implications of a dilaton in 4d gauge
theories are revisited. In part I of this paper we see how an
effective dilaton coupling to gauge kinetic term provides a simple
attractive mechanism to generate confinement. In particular, we
put emphasis on the derivation of confining analytical solutions
and look into the problem how dilaton degrees of freedom modify
Coulom potential and when a confining phase occurs. In part II, we
solve the semi-relativistic wave equation, for Dick interquark
potential using the Shifted l-expansion technique (SLET) in the
heavy quarkonium sector. The results of this phenomenological
analysis proves that these effective theories can be relevant to
model quark confinement and may shed some light on confinement
mechanism. \keywords{dilaton, confinement, quark potential}

\end{abstract}

\section{Introduction}
\label{sec:intro}
Despite enormous amount of work performed over
more than thirty years, particularly in lattice simulations of
QCD, full Understanding of the QCD vacuum structure and color
confinement mechanism are still lacking. Indeed, direct derivation
of confinement from first principles remain still elusive and
there is no totally convincing proposal about its generating
mechanism. However, on the other hand, we known that the vacuum
topological structure of theories with dilaton fields is
drastically changed compared to the non dilatonic ones \cite{CT}.
Therefore much about confinement might be learned from such
theories, particularly string inspired ones. The presence of
fundamental scalars with direct coupling to gauge curvature terms
in string theories offers a challenge with attractive implications
in four-dimensional gauge theories.  \footnote{The dilaton is an
hypothetical scalar particle predicted by string theory and
  Kaluza-Klein type theories. Its expectation
  value  probes  the strength of the gauge
coupling \cite{GSW}.} Besides, since color confinement can be
signaled through the behavior of the interaction potential at
large distances. In this context, it was suggested in \cite{dick}
that an effective coupling of a massive dilaton to the
4-dimensional gauge fields may provide an interesting mechanism
which accommodates both the Coulomb and confining phases. The
derivation performed in \cite{dick,Ch1} suggest a new scenario to
generate color confinement. This scenario may be considered as a
challenge to the mechanism based on monopole condensation.

The outline of this paper is as follows. In the part I, we
describe the influence of the dilaton on a low energy gauge theory
and look into the problem how dilatonic degrees of freedom
modifies Coulomb potential and how transition to a confining phase
occurs. Then, we review several recent work by presenting the
corresponding effective coupling functions used. We briefly
comment on the analytic solutions of the field equations and their
confinement features. Part II is devoted to phenomenological
investigations. We study Dick interquark potential in the heavy
quarkonium systems using SLET technique. and summarize the results
obtained from this analysis. Finally, we draw our general
conclusion.

\section{The low energy effective theory}

The imprint of dilaton on a 4d effective nonabelian gauge theory
is described by a Lagrangian density:

\begin{equation}
{\cal L}({\phi},A)=
-\frac{1}{4F({\phi})}{G_{{\mu}{\nu}}^a}{G^{{\mu}{\nu}}_a}
+\frac{1}{2}\partial_\mu \phi \partial^\mu \phi  -V(\phi) +J_a^\mu
A_\mu^a
\end{equation}

 where $\phi$ is the dilaton field and $G^{\mu \nu}$ is the standard field strength tensor of the
theory.  $V(\phi)$ denotes the non perturbative dilaton potential
and  $F(\phi)$ represents the coupling function depending on
$\phi$. Several forms of $F(\phi)$ have been proposed in
literature. The most popular one $F(\phi)=e^{-k\frac{\phi}{f}}$
occurred in string theory and Kaluza-Klein theories \cite{GSW}.

Analysis of the problem of  Coulomb gauge theory augmented
with dilaton degrees of freedom in (1) performed as follows:\\
First, we consider a point like static Coulomb source defined in
the rest frame by the current:

\begin{equation}
J_a^\mu =g \delta (r) C_a \nu_0^\mu =\rho_a \eta_0^\mu
\end{equation}
where $C_a$ is the expectation value of $SU(N_c)$ generator. The
field equations emerging from the static configuration (2) are
given by:

 \begin{equation}
 \left[ D_\mu , F^{-1} (\phi ) G^{\mu\nu}\right] = J^\nu
\end{equation}

 and

\begin{equation}
 \partial_\mu \partial^\mu \phi = -\frac{\partial
 V(\phi)}{\partial\phi}-\frac{1}{4} \frac{\partial F^{-1}(\phi)}{\partial
 \phi}G_a^{\mu\nu}G_a^{\nu\mu}
 \end{equation}

By setting $G_a^{0i} = E^i \chi_a =-\nabla^i \Phi_a$, and after
some algebra, we derive the chromo-electric field:

\begin{equation}
E_a=\frac{Q^a_{eff}(r)}{r^2}
\end{equation}

where the effective charge is $$Q^a_{eff}(r)=\left(g\frac{C_a}{4\pi}\right) F(\phi(r))$$. \\
Eq(5) shows that it is the running of the effective charge that
makes the potential stronger than the Coulomb potential. In other
words, the Coulomb spectrum is recovered if the effective charge
did not run. Thereby the interquark potential reads as \cite{Ch1},

\begin{equation}
U(r)=2\widetilde{\alpha}_s \int \frac{F(\phi(r))}{r^2} dr
\end{equation}

with $\alpha_s = \frac{g^2}{4\, \pi}$ and $\widetilde{\alpha}
=\frac{\alpha_s}{8\pi} \left( \frac{N_c -1}{2N_c}\right)$

The formula in Eq. (6) is remarkable since it provides a direct
relation between the interquark potential and the coupling
function $F(\phi(r))$. Moreover, it shows that existence of
confining phases in this effectivee theory is subject to the
following condition,

\begin{equation}
\lim_{r\to \infty} r F^{-1}(\phi(r)) = finite
\end{equation}

At this stage, the main objective is to solve the field equations
of motion (3) and (4) and determine analytically $\phi(r)$ and
$\Phi_a(r)$. For this, $F(\phi)$ and $V(\phi)$ have to be fixed.
In the sequel the dilaton potential is set to $V(\phi)
=\frac{1}{2} m^2 \phi$. Below, we will briefly describe the main
features of three recent models and present their solutions.


\subsection{ Dick Model}

In this effective theory, Dick used the form: $\frac{1}{F(\phi)} =
\frac{\phi^2}{f^2 + \beta\, {\phi}^2}$ where f represents a
coupling scale characterizing the strength of the scalar-gluon
coupling and $\beta$ is a parameter in the range $\Big[0, 1\Big]$.
He derived the radial dependence of the dilaton field and the
interquark potential (up to a color factor) \cite{dick}:
$$ \phi(r)={\pm}\frac{1}{r}\sqrt{\frac{k}{m}+({y_0^2}-\frac{k}{m})exp(-2mr)}$$

$$ V(r) = [\frac{\beta\,  g^2}{4\, \pi\, r}
-gf  \sqrt  \frac{N_c}{2\,(N_c - 1)} ln[e^{2mr} - 1 +
\frac{m}{k}y_o^2]$$

With the abbreviation: $k^2=\frac{\alpha_s\, f^2}{8\,
\pi}\frac{N_c-1}{N_c}
$ \\

Remarkably the potential $V(r)$ comes with the required behavior:
a first term which accommodates the Coulomb interaction at short
distances and a second term linearly increasing in the asymptotic
regime with a string tension \footnote{In the massless case, $
V(\phi)=0$, solutions of the field equations reduced to: $\phi(r)
= {\pm}\big(\frac{g\,f}{2\pi}\big) \, \sqrt{\frac{N_c -
1}{N_c}}\,r^\frac{-1}{2}$, $V(r) = \frac{g^2\, \beta\, (N_c -
1)}{8\,\pi\,r N_c} -
\frac{f\,g}{2}\sqrt{\frac{N_c - 1}{N_c}}\,r $} \\

$\sigma  \sim g\, m\, f$ which depends on the dilatonic degrees of freedom $m$, $f$.\\

\subsection{ Cornwall-Soni Model}

In this model, the glueballs are represented by a massive scalar
field $\phi$ and couple in a non minimal way to gluons, through $
\frac{1}{F(\phi)} =
\frac{\phi}{f}$ \cite{CS}. Cornwall-Soni were the first to motivate such term as a low enrgy correction to effective models of QCD\\

Analytical Solutions were found for $r  \to \infty $\cite{DF},

$$  \phi(r) =  \Big[\frac{\alpha_s\, f \, (N_c - 1)}{16\, \pi\, m^2\, N_c}\Big]^{\frac{1}{3}} \,
  r^{-\frac{4}{3}} $$
$$V(r) = -3 \, g \frac{N_c - 1}{2\, N_c} \, \Big[ \frac{g \, f^2 \, N_c
\, m^2}{\pi \, (N_c - 1)}\Big]^\frac{1}{3}r^{\frac{1}{3}}$$

These formulas show that  at large distances, confinement is
probed through an interaction potential proportional to $r^{1/3}$
and considered by the authors as non perturbative correction to
the Coulomb phase.

\subsection{ Chabab-Sanhaji Model}

The main aim in this work was to construct a low energy effective
field theory from
which some popular phenomenological potentials may emerge. To this end,  we proposed the coupling function $F(\phi)=\Big( 1 -\beta \frac{\phi^2}{f^2}\Big)^{-n}$ \cite{chababsanhaji}.\\
By substituting $F(\phi)$ Eq. (3, 4), the field equations were
found too complicated to integrate analytically. Fortunatly,
since the focus is on the long range behavior of the dilaton field
and on how it modifies the Coulom phase, the analysis is
restricted to the infrared region. Thus, the asymptotic solutions
are found to be,

$$\phi=\Big[ \frac{f^2}{\beta}-\Big(\frac{\beta}{f^2}\Big)^\frac{-n}{n+1}\Big( \frac{2n\alpha_s}{m^2}\Big)^\frac{1}{n+1}\Big(\frac{1}{r}\Big)^\frac{4}{n+1}\Big] $$
and the chromo-electric potential:
$$\Phi_a(r)=-\frac{gC_a}{4\pi}\Big(\frac{2n\alpha_s}{m^2f^2} \Big)^\frac{-4n}{n+1}\frac{n+1}{3n-1}r^{\Big(\frac{3n-1}{n+1}\Big)} $$

We see that the occurrence of confinement depends on the parametre
$n$ and our effective theory can serve to model quark confinement
when $n\in \Big[\frac{1}{3}, 1\Big]$.

On the other hand, going back to the objective of this study: by
selecting specific values of $n$, we reproduced the following
known interquark potentials
\begin{itemize}
\item   $n=1$ $\Rightarrow$  linear term of Cornwall potential \cite{Ei}.
\item   $n=11/29$ $\Rightarrow$ Martin's potential \cite{Martin}.
\item   $n=3/5$  $\Rightarrow$ Song-Lin, or Motyka-Zalewski' potential \cite{motyka}.
\item   $n=5/9$ $\Rightarrow$ Turin potential \cite{Turin}.
\end{itemize}

Therefore, these quark potentials, which gained credibility only
through confrontation to the hadron spectrum, are now supplied
with a theoretical framework since they can be derived from a low
energy effective theory.


\section{ Phenomenological Analysis}

Our aim in this part of the review paper is to dedicate more
efforts to understand the new confinement mechanism suggested
above through phenomenological investigation of Dick potential in
the heavy meson sector. This study will be addressed as in
\cite{barakat} where the shifted-$l$ expansion technique is used
(SLET) where $l$ is the angular momentum. This method  provides a
powerful analytic technique for determining the bound states of
the semi-relativistic wave equation consisting of two quarks of
masses $m_{1}$, $m_{2}$ and total binding meson energy $M$ in any
spherically symmetric potential. It is rapidly converging and
handles highly excited states which pose problems for variational
methods \cite{sung}. Moreover, relativistic corrections are
included in a consistent way. \\

 Dick interquark potential reads,
\begin{equation}
V_D(r)=-{4 \over 3}{\alpha_s \over r}+{4 \over
3}gf\sqrt{N_c\over{2(N_c-1)}}\ln[exp(2mr)-1]
\end{equation}

The SLET technique used to obtain results from the theory requires
us to
 specify several inputs: $m_{c}$, $m_{b}$, $m$, $f$ and
$\alpha_s$. In our numerical analysis, we set the charm and bottom
quark
 masses to the values $m_{c}=1.89$~GeV and $m_{b}=5.19$~GeV. For the QCD
 coupling constant, in contrast to the Lattice potentials which use the same
 effective coupling in the description of heavy quarkonium, we take into account the running of $\alpha_s$,
\begin{equation}
\alpha_s(\lambda)=\frac{\alpha_s(m_z)}
{1-(11-\frac{2}{3}n_f)[\alpha_s(m_z)/2\pi]ln(m_z/\lambda)},
\end{equation}
 where the renormalization scale is fixed to  $\lambda=2 \mu$, with $\mu$ is the reduced mass,
\begin{equation}
\mu=\frac{m_1m_2}{m_1+m_2},
\end{equation}

Thus, combination of the leading order formula (9) and the world
experimental value $\alpha_s(m_z)=0.12$ yields,
\begin{equation}
\alpha_s (charmonium)=0.31,  \qquad \alpha_s (bottomonium)=0.20 ,
\qquad
\end{equation}
\noindent  while $\alpha_s=0.22$ for the $b\bar{c}$ quarkonia. On
the other hand, the interquark potential parameters  $m$ and $f$
are treated as being free in our analysis and are obtained by
fitting the spin-averaged $c\bar{c}$ and $b\bar{b}$ bound states.
An excellent fit with the available experimental data can be seen
to emerge when the following values are assigned \footnote{if the
standard value for the string tension $0.18~ GeV^2$ is used, the
dilaton mass will be shifted to a value about $158~ MeV$.},
\begin{equation}
m=57~ MeV \qquad  gf \sqrt {\frac{N_c}{2(N_c-1)}}=430~ MeV.
\end{equation}

\begin{table}[ht]
\tbl{Calculated mass spectra (in units of GeV) $M_{n\ell}$ of
  $c\bar{c}$ boundstates from  Dick interquark potential}
{\begin{tabular}{|c|c|c||c|c|c|} \hline
 State,$n\ell$ &$M_{n\ell}$, SLET &$M_{n\ell}$, Exp.&
State,$n\ell$ &$M_{n\ell}$, SLET &$M_{n\ell}$, Exp. \\ \colrule
1S &3.073  &3.068  &1P&3.546 &3.525\\
2S & 3.662 &3.663  &2P&3.871 &-  \\
3S &4.027  &4.028  &1D&3.787&3.788   \\
\hline
\end{tabular}  \label{Table1}}
\end{table}
\begin{table}[ph]
\tbl{Calculated mass spectra (in units of GeV) $M_{n\ell}$ of
$b\bar{b}$ from Dick interquark potential}
{\begin{tabular}{|c|c|c||c|c|c|} \hline
 State,$n\ell$ &$M_{n\ell}$, SLET &$M_{n\ell}$, Exp.&
State,$n\ell$ &$M_{n\ell}$, SLET &$M_{n\ell}$, Exp. \\ \colrule
1S &9.450 &9.446  &1P&9.903 &9.900\\
2S & 10.014  &10.013 &2P &10.227 &10.260  \\
3S &10.299  &10.348 &1D &10.129&-   \\
\hline
\end{tabular}  \label{Table2}}
\end{table}
\begin{table}[ph]
\tbl{Calculated mass spectra (in units of GeV) $M_{n\ell}$ of
$b\bar{c}$ boundstates from Dick potential}
{\begin{tabular}{|c|c|c||c|c|c|}   \hline
 State,$n\ell$ &$M_{n\ell}$, SLET &$M_{n\ell}$, Exp.&
State,$n\ell$ &$M_{n\ell}$, SLET &$M_{n\ell}$, Exp. \\  \colrule
1S &6.322 &6.276 $\pm0.004$$\pm0.0027$ &1P &6.767&-\\
2S & 6.876 &- &2P&7.072&-  \\
3S&7.181 &- &1D&6.994&-   \\
\hline
\end{tabular}  \label{Table3}}
\end{table}

Tables (1,2) list the results of the analysis for the
spin-averaged energy levels of interest. In all cases, where
comparison with experiment is possible, agreement is generally
very good. Next step, to check the consistency of our predictions,
we  estimate the bound states energies of the $b\bar{c}$
quarkonia. These states are  expected to be produced at LHC and
Tevatron. Moreover, they should provide an excellent test to
discriminate between various techniques used to probe
nonperturbative properties of hadrons. In table 3 we show our
calculated spectrum. The estimate of the  mass of the lowest
pseudoscalar S-state of the $B_c$ spectra is close to the
experimental value reported by CDF and D0 collaborations
\cite{cdf}. As to the higher states masses, they compare favorably
with other predictions based on QCD sum-rules \cite{kis2,chab3} or
potential models \cite{quigg}-\cite{brambilla}. In conclusion,
Dick interquark potential (08) is tested successfully to fit the
spin-averaged  quarkonium spectrun. In view of these results, it
is quite encouraging to pursue phenomenological application of
$V_D(r)$ and other quark potentials emerging from such low
effective gauge theory with dilaton.

\section{General conclusion}

We revisited some of the most recent work on confinement in 4d non
abelian gauge theories with a massive scalar field (dilaton) and
effective coupling functions to gauge fields.  Analytical
solutions have been found with confinement feature at large
distances. Thus, These low energy effective theories can serve
well to model quark confinement. Moreover, by using Dick
interquark potential in the heavy quarkonium sector, we showed
that phenomenological investigation of the confinement generating
mechanism suggested by these models is more than justified.
Indeed, the obtained results for charmoniun and bottomonium fit
well experimental data when the dilaton mass is given a value
about 57 MeV. Also, for $B_c$ system, we found that the S-state
energy level is close to the value reported by CDF collaboration,
while those of excited states agree favorably with  predictions of
other theoretical studies. On the other hand, as a by-product,
this analysis allows a test to the physics beyond the standard
model in relation to hadron spectroscopy. Indeed, the estimate of
the dilaton mass lies in the range of values proposed in
\cite{gasperini,cho}. which may shed some light on the search of
the dilaton since the possibility to identify this hypothetical
particle to a fundamental scalar invisible to present day
experiments should not be ruled out
 \cite{bando,halyo}.

\section*{Acknowledgments}
This Work is partially supported by the Morocco research program
PROTARS III, under contract D16/04.


\begin{thebibliography}{0}


\bibitem{CT}
M. Cvetic, A.A. Tseytlin, {\it Nucl. Phys.} {\bf B 416}, 137
(1983).
\bibitem{GSW}
M. Green, J. Schwartz, E. Witten, Superstring Theory, (Cambridge
University Press, Cambridge 1987)
\bibitem {dick} R. Dick {\it Eur. Phys. J.} {\bf C6 } 701 (1999); {\it Phys. lett.
} {\bf B 397}, 193 (1997); {\it Phys. lett.} {\bf  B 409}, 321
(1997).
\bibitem{Ch1}
M. Chabab, R. Markazi and E. H. Saidi, {\it Eur. Phys. J.} {\bf C
13}, 543 (2000).
\bibitem{CS}
J. M. Cornwall and  A. Soni, {\it Phys. Rev.} {\bf D 29}, 1424
(1984).
\bibitem{DF}
R. Dick and L. P. Fulcher, {\it Eur. Phys. J.} {\bf C 9}, 271
(1999).
\bibitem{chababsanhaji}
M. Chabab and L. Sanhaji,  {\it Int. J. Mod. Phys.} {\bf A 20}
1863 (2005);  {\bf hep-th/0311096}.
\bibitem{Ei}
E. Eichten et al.,{\it Phys. Rev. Lett.} {\bf 34}, 369 (1975).
\bibitem{Martin} A. Martin, {\it Phys. Lett.} {\bf B 100} 511 (1981).
\bibitem{motyka} L. Motyka and K. Zalewski,  {\it Z. Phys.} {\bf C 69} 342
  (1996); X. Song and H. Lin, {\it Z. Phys.} {\bf C 34}, 223 (1987).
\bibitem{Turin} D.B. Lichtenberg et al., {\it Z. Phys.} {\bf C 41} 615 (1989)
 107.
\bibitem {barakat} T. Barakat, {\it Int. J. Mod. Phys.} {\bf A 16} 2195 (2001).
\bibitem {sung} D. Sung Hwang and G. Hee Kim, {\it Phys. Rev.} {\bf D 53}
 3659 (1996); D. Sung Hwang et al., {\it Phys. Rev.} {\bf D 53}, 4951 (1996).
\bibitem{cdf} F. Albe et al. (CDF Collaboration), {\it Phys. Rev. Lett.} {\bf 81}
  2432 (1998); R.K. Mommeson (on behalf of CDF and D0 collaborations){\bf hep-ex/0612003}.
\bibitem{kis2} V.V. Kiselev, {\it Int. J. Mod. Phys.} {\bf A 11} 3689 (1996).
\bibitem{chab3} M. Chabab, {\it Phys. Lett.} {\bf B 325} 205 (1994); {\it 7th
   Inter. Conf. Hadon Spectroscopy}, AIP
  Conference Proceedings {\bf 432}, Upton, New York, 1997, 856.
\bibitem{quigg} E. Eichten and C. Quigg, {\it  Phys. Rev.} {\bf D 49} 584 (1994).
\bibitem{gersh} S.S. Gershtein, V.V Kiselev, A.K. Likhoded and A.V. Tkabladze,
  {\it Phys. Rev.} {\bf D 51},  3613 (1995).
\bibitem{godfrey} S. Godfrey and N. Isgur, {\it Phys. Rev.} {\bf D 32}, 189
  (1985); S. Godfrey, {\it Phys. Rev.} {\bf D 70}, 054017 (2004).
\bibitem{zhang} J. Zhang, J.W. Van Orden and W. Roberts, {\it Phys. Rev.} {\bf D
    52}, 5229 (1995).
\bibitem{gupta} S.N. Gupta and J.N. Johnson, {\it Phys. Rev.} {\bf D 53}, 074006
  (1996).
\bibitem{ebert} D. Ebert, R.N. Faustov and V.O. Galkin, {\it Phys. Rev.} {\bf D
    67}, 014027 (2003).
\bibitem{brambilla} N. Brambilla et al., {\it CERN Yellow Report on Heavy
    Quarkonium}, {\it hep-ph/0412158}  (2004).

\bibitem{gasperini} M. Gasperini,  {\it Phys. Lett.} {\bf B 327} 214 (1994).
\bibitem{cho} Y.M. Cho and Y.Y Keum, {\it Mod. Phys. Lett.} {\bf A 3} 108 (1998).
\bibitem{bando} M. Bando, K. Matumoto  and K. Yamawaki,  {\it Phys. Lett.} {\bf
    B 178}, 308 (1986).
\bibitem{halyo} E. Halyo , {\it Phys. Lett.} {\bf B 271}, 415 (1991).

\end{thebibliography}
\end{document}